\documentclass[11pt]{article}
\usepackage{epsfig}

\newcommand{\ca}{{\cal A} }
\newcommand{\bq}{{\overline Q} }

\newcommand{\be}{\begin{equation}}
\newcommand{\ee}{\end{equation}}
\newcommand{\ba}{\begin{array}}
\newcommand{\ea}{\end{array}}
\newcommand{\bea}{\begin{eqnarray}}
\newcommand{\eea}{\end{eqnarray}}
\newcommand{\bc}{\begin{center}}
\newcommand{\ec}{\end{center}}
\newcommand{\bi}{\begin{itemize}}
\newcommand{\ei}{\end{itemize}}

\newcommand{\disregard}[1]{{}}

\def\bild#1\over#2{\mathrel{\mathop{\kern0pt #1}\limits_{#2}}}

\begin{document}
\ 
\vskip 2cm
\centerline {\large{\bf FIELD THEORETICAL COMPUTATIONS OF STATIC CONDUCTIVITY}}
\centerline{\large {\bf FOR A WHITE NOISE MAGNETIC SYSTEM\rm}}
\vskip 2cm                 
\centerline{Cyril FURTLEHNER\rm \footnote{cyril.furtlehner@fys.uio.no}} 
\hfill\break
\centerline{\it Institute of Physics, University of Oslo}
\centerline{\it P.O. Box 1048 Blindern}
\centerline{\it N-0316 Oslo, Norway}
\vskip 2cm
\hfill\break
\centerline{\large\bf Abstract} 
\hfill\break

{\large We  consider an  electron coupled to a random magnetic 
  field with local correlations and eventually non-zero average value.
Starting from the source-term formalism of Levine, Libby and Pruisken, we define a generating function for
static transport coefficients. We use  a non-hermitian representation of the Hamiltonian, and a functionnal 
representation of the generating function, where the integration variable consists of non-local supermatrices,
due to the non-locality of the Aharanov-Bohm interaction.
The conductivity coefficients  are computed using a specific saddle-point procedure, which incorporates 
ladder diagrams contributions. 
The situation with an external uniform magnetic field is considered  with the two different spin projection possibilities. The limit 
of vanishingly  small external magnetic field is unstable with respect to spin flip, because of $E=0$ states.
In particular a quantized step is obtained   when comparing the 
Hall conductance between the two possible spin projections.}

\vfill\eject

\hfill\break
{\bf\large 1. Introduction}
\hfill\break

The problem of an electron moving in a random magnetic field  has been in the recent years
subject of a great interest in particular because of its relevance to the Hall problem, especially at half filling \cite{HLR}. 
It may be defined in many ways, either on a Lattice with random fluxes \cite{IFK,BFZ} or in the continuum with various choices
for the statistical correlations of the magnetic field, by considering for example random distribution of point-like
vortices \cite{DFO1,DFO2}.
The spectrum properties of these models are now well established.
Although weak-localization has 
been studied \cite{RAS}, diffusion modes have been identified, and an effective field theory for weak disorder has been 
constructed \cite{AMW},  
the question of localization is however still under debate. Most recent results seem to converge to the idea that all states are
localized except at $E=0$ for lattice models \cite{ZHA,ASM}.
The question of transport coefficient has been less studied. As noticed by Lusakowski and Turski \cite{LTI2},
there might be some additionnal
features due to the random nature of the velocity operator. Concerning this aspect, some qualitative and quantitative perturbative
results have been obtained in the situation of random magnetic impurities and strong uniform magnetic field, in particular oscillations
of the Hall conductance \cite{DOT,EVB}.

In the present article, we are interested in using some field theoretical methods, originally devised in the context 
of localization, to consider this question of transport coefficients. Especially we would like, as a first step,
to compute the SCBA approximation
in a well defined manner, using this field theoretical language. We will consider a two dimensional model were the magnetic fluctuations 
are distributed  with uncorrelated gaussian distribution and are added  to an eventual  uniform magnetic field.
In the first part, as a preliminary, we use the source term formalism
of Levine, Libby and Pruisken to define a generating function for static transport coefficients, which can be seen as a 
vacuum to vacuum transition amplitude at the Fermi energy,
and necessitate to consider sources in matrix form. As a byproduct it is a generating function
for Ward identities. In the second part, the model under
consideration is specified, and a non-unitary transformation is performed, which leaves the generating function invariant
and simplifies the averaging procedure. The generating function is then expressed, using functionnal integration over
non-local replicated supermatrices, in consequence of the non-locality of the Aharonov-Bohm interaction. This is the main technical 
difference with the standard random scalar potential. The saddle-point is then exactly solved and leads to semi-circle type expression
for the DOS. For the issue of transport coefficients, the saddle point equation is then reconsidered in presence of the sources, which 
define in an unambiguous manner the saddle-point expression for these transport coefficient, and allow to eventually 
incorporate ladder diagrams contributions, which are known to be exclusively responsible of IR divergencies in absence of time reversal
symmetry. Some perspective and natural extensions of this work,
will then be drawn  in the conclusive section.

\hfill\break   
{\bf\large 2. Generating function for static transport coefficients}\\
{\it\large 2.1 Smrcka-Streda formula}
\hfill\break

We start with the expressions for the static conductivity coefficients, obtained from the 
linear response theory. Given the general form of the Hamiltonian
\be\label{H} H={1\over2m}({\bf p}-e{\bf A})^2\ +\ V({\bf r})\ee
and assuming a zero temperature Fermi-Dirac equilibrium  distribution, longitudinal and transverse transport coefficients 
can be expressed as Fermi-energy quantities, in the following way ($\hbar=1$) \cite{SMS}
\be \sigma_{xx}={\pi e^2\over V}{Tr}[v_x\delta(E_F-H)v_x\delta(E_F-H)]\ee
and 
\bea \sigma_{xy}&=&\sigma_{xy}^I+\sigma_{xy}^{II}\\
    \sigma_{xy}^{I}&=&{i e^2\over2 V}{\bf Tr}[v_x G^{+}v_y\delta(E_F-H)-v_x\delta(E_F-H)G^{-}v_y]\\
    \sigma_{xy}^{II}&=&{e^2\over2 V}{\bf Tr}[(xv_y-yv_x)\delta(E_F-H)]\eea
$G^{\pm}=(E_F-H\pm i\eta)^{-1}$ are the advanced (resp. retarded) Green function.
$v_i=p_i-eA_i$ are the components of the velocity operator.
$V$ is the volume, and these expressions represent an average over the sample 
of local current-current correlation function, concerning $\sigma_{xx}$ and $\sigma_{xy}^{I}$, although
$\sigma_{xy}^{II}$ is the magnetization at the Fermi energy per unit volume of the sample. It is already to be noticed that this 
last formula is not well defined unless edges are taken into account \cite{HLP,PSK}. A way to avoid a cumbersome regularization 
is to use the Streda formula \cite{STD}
\be\label{std}  \sigma_{xy}^{II}=e{\partial N(E)\over\partial B}\Big\vert_{E=E_F}\ee
where $N(E)$ is the integrated density of states.
Since the model we have in mind is two dimensional, it will be convenient to use complexe notations ($z=x+iy$, $A_z=A_x+iA_y$, 
$\partial_z={1\over2}(\partial_x-i\partial_y)$, 
$v_z={1\over2}(v_x-iv_y)$).
Using these notations, we can recast the coefficients in the following manner
\bea\label{LRC}\sigma_{xx}&=&{2\pi e^2\over V}{\bf Tr}[v_z\delta(E_F-H)v_{\bar z}\delta(E_F-H)]\\
    \sigma_{xy}^{I}&=&{i e^2\over\pi V}{\bf Tr}[v_z G^{-}v_{\bar z}G^{+}-v_zG^{+}v_{\bar z}G^{-}]\\
    \sigma_{xy}^{II}&=&{ie^2\over2 V}{\bf Tr}[(zv_z-\bar zv_{\bar z})\delta(E_F-H)]\eea
In the expression of $\sigma_{xy}^{I}$, has been used  the relation 
\be\label{GWI} {\bf Tr}v_zG^{\pm}v_{\bar z}G^{\pm}=-{1\over2}{\bf Tr}G^{\pm}\ee
a direct consequence of the commutation rules
\bea\label{CR} v_z&=&{i\over2}[\bar z,G^{-1}]\\
     v_{\bar z}&=&{i\over2}[z,G^{-1}]\eea
\hfill\break
{\it\large 2.2 Generating function}
\hfill\break
 
The Field theoretic approach to localization led to realize \cite{WGR} the intimate relation between d.c. transport (at $T=0$)
and the symmetry between advanced and retarded Green function; the localization regime corresponds to the restauration
by quantum interferences
of this symmetry \cite{MKS,EFT}, otherwise spontaneously broken by a non-vanishing density of states at the Fermi energy.
An explicit illustration of this, is the source-term formalism of Levine, Libby and Pruisken \cite{LLP}, which 
expresses the d.c. transport coefficients as the response of the system to some rotation between the advanced and retarded wave.
By convenience we will reformulate slightly this formalism in the following.

Consider the generating function
\be {\cal F}[\ca_z^{\alpha},\ca_{\bar z}^{\alpha}]=\log\ {\det[E_F-H(\ca_z,\ca_{\bar z})+i\eta\tau_3]\over
                                                    \det[E_F-H+i\eta\tau_3]}\ee
It is  left implicit here, that the operators involved in this expression acts on two-components wave functions, 
corresponding to either the retarded or advanced sector, as emphasized by the presence of the infinitesimal prescription
$i\eta\tau_3$.
The $\ca_z^{\alpha}$ are the components of  the sources $\ca_z=\ca_z^{\alpha}\tau_\alpha$, which are
of matrix form, $\tau_\alpha$ beeing the usual Pauli matrices
\be     \tau_1=\pmatrix{0&1\cr 1&0\cr}\qquad\tau_2=\pmatrix{0&-i\cr i&0\cr}\qquad\tau_3=\pmatrix{1&0\cr 0&-1\cr}\ee
In addition, we have
\be H(\ca_z,\ca_{\bar z})=\left\{(v_z-{e\over2}\ca_{\bar z}),(v_{\bar z}-{e\over2}\ca_z)\right\}
                          \ +\ V(z,\bar z)\ee
(where $\{,\}$ means anticommutator).
If we restrict now the sources to be spatially uniform for $\alpha=1,2$ and 
\be \ca_z^3={iB^3\over2}z\ee
$B^3$ being a uniform magnetic field
then, the static transport coefficients can be expressed in the following way
\bea \sigma_{xx}&=&-{1\over2\pi V}\ {\partial^2{\cal F}\over\partial\ca_z^1\partial\ca_{\bar z}^1}\Big\vert_{\ca=0}\\
    \sigma_{xy}^{I}&=&-{1\over\pi V}\ {\partial^2{\cal F}\over\partial\ca_z^1\partial\ca_{\bar z}^2}\Big\vert_{\ca=0}\\
    \sigma_{xy}^{II}&=&-{1\over2i\pi V}\ {\partial{\cal F}\over\partial B^3}\Big\vert_{\ca=0}\eea
by direct comparison with (\ref{LRC},8,9), using the expansion
\be {\cal F}[\ca_z^{\alpha},\ca_{\bar z}^{\alpha}]=\log\det[1-\Delta H G]=-{\bf Tr}[\Delta H G+{1\over2}(\Delta H G)^2+...]\ee
with $\Delta H=H(\ca_z,\ca_{\bar z})-H$ and $G=\pmatrix{G^+&0\cr 0&G^-\cr}$, and using the Ward identity (\ref{GWI}) 
which in fact is a consequence of global $U(1)\times U(1)$ invariance of $\cal F$ \cite{LLP}.
The relation concerning $\sigma_{xy}^{II}$ is reminiscent of the Streda formula (\ref{std}). The connection is simply made 
by realizing that
\be {\partial{\cal F}[\ca_z^{3},\ca_{\bar z}^{3}]\over\partial E_F}={\bf Tr}[G^+(B+B^3)+G^-(B-B^3)-G^+(B)-G^-(B)]\ee
(where the argument of the $G$'s denotes the sum of the magnetic field $B$ and the fictious one coming from the source $\ca^3$).
Since
\be G^{\pm}(B)= Re(B)\mp i\pi\rho(E_F,B)\ee
($Re(B)$ is the real part of the Green function, and $\rho(E_F,B)$ is the density of states at the Fermi energy and for a given 
magnetic field $B$). Taking the first order term in the last relation, and integrating over energy leads to the relation
\be {\partial{\cal F}\over\partial B^3}=-2i\pi V{\partial N(E_F)\over\partial B}\ee
which makes the connection with the Streda formula.
The underlying reasons for these relations concerning $\sigma_{xy}^{II}$
 is that the source $\ca^3$ is the conjugate field of the persistent current,
locally  given (at position ${\bf z}=(z,\bar z)$) by
\be j_z^0({\bf z})={i\over2\pi}{\bf Tr}\left[v_z\vert{\bf z}><{\bf z}\vert\ G\ \tau_3\right]\ee
This entails, that if $\ca^3$ corresponds to a magnetic field, 
\be B^3={1\over i}(\partial_z\ca_z^3-\partial_{\bar z}\ca_{\bar z}^3)\ee
then this magnetic field is conjugate to the magnetization (at the Fermi-energy), locally defined by
\be M({\bf r})={1\over2i\pi}
\int d^2{\bf z}\left({1\over \bar z-\bar z'}j_z^0({\bf z}')-{1\over z-z'}j_{\bar z}^0({\bf z}')\right)\ee
meaning that this quantity can be obtain from the generating function in the folowing manner
\be M({\bf r})=-{1\over2i\pi}{\partial{\cal F}\over\partial B^3({\bf r})}\Big\vert_{\ca=0}\ee
This is quite similar also to formulaes derived in the context of scattering theory \cite{AKA,FKR,MRZ} where the logarithm
of the determinant of the scattering matrix is used as a generating function instead of $\cal F$.\\ 
\hfill\break
{\bf\large 3. The model}\\
{\it\large 3.1 Pauli equation}
\hfill\break

After this preamble concerning the transport coefficients, we are now in position to define the model 
we are interested in, namely to specify equation (\ref{H}). As already mentioned in the introduction, the random magnetic problem
should be relevant for studying transport in a fractionnal quantum-Hall state, especially at even denominator filling factors. 
In this case, the ground-state consists of an assembly of composite Fermions,
forming a weekly interacting  Fermi-sea, and scattered by randomly distributed impurities. Due to the Chern-Simons gauge field
needed to describe the 
ground-state, the effect of impurities is to produce fluctuations in the fictious magnetic field  by modifying the local
density of particules \cite{HLR} . This entails the presence of magnetic vortices, located on the top of these impurities.
For example, in a $\nu={1\over q}$ fractionnal state, a delta impurity is expected to be accompagned with $q$ quantum of flux.
As a consequence, a relevant model for this problem could actually be defined by
\be\label{HLA} H={1\over2m}({\bf p}-e{\bf A})^2\ +\lambda B({\bf r})\ee
where $B({\bf r})$ is a random magnetic field plus the eventual effective
external field, ${\bf A}$ the corresponding static gauge field, 
and $\lambda$ a proportionality
constant between the impurity scalar potential an the corresponding induced magnetic field.
Once the statistical properties of the magnetic field are specified, the model is entirely defined.
Instead of studying this model for an arbitrary value of the parameter $\lambda$, we will restrict ourselves to the two
particular values $\lambda=\pm {e\over2 m}$. In the context of the Aharonov-Bohm problem  \cite{BLZ,OVY},
or Chern-Simons theory \cite{JWP} these  particular 
values correspond to the situation where the scale-invariance of the theory is not broken quantum mechanically 
and can be viewed as  fixed point of the parameter $\lambda$ under renormalization.
They are moreover interpreted to be 
the spin-coupling constant of the electron with the magnetic field \cite{CMO}, the equation (\ref{HLA})
beeing  then precisely the Pauli equation, with the sign depending on the projection of the spin along the $z$ axis.
In the present context, this particular choice has as well the virtue
of eliminating ultraviolet  divergences, which otherwise occur at arbitrary $\lambda$ in the perturbation expansion (of the average 
density of states for example), indicating also the scale invariance of the theory for these particular values. As a consequence
we will consider this to be the most relevant situation, and postpone
the analyses for arbitrary $\lambda$ to further investigations.\\ 
For simplicity, we will consider a locally correlated random magnetic field, in its most simple expression, namely the white noise
case,
\be \overline{B({\bf r})B({\bf r}')}= w \delta({\bf r}-{\bf r}')\ee
and with all the higher order cumulants set to zero. If we compare to a poissonian distribution of $\delta$ fluxes $\phi$, with 
mean density n, the white noise distribution amounts to conserve only the second cumulant with a coefficient equal to
\be  w=n\phi^2\ee
This approximation, to some extent, is justified from  the field theoretical point of view, for scalar impurities (not magnetic).
The second cumulant gives rise to a local quartic interaction term, and other cumulants contribute
to terms of higher degre in the interaction, which are in principle irrelevant. This might be less justified in the random magnetic-field
problem (see reference \cite{DFO1} for the average density of states for poissonian magnetic impurities,
\cite{DFO2} for non-locally  
correlated poissonian magnetic impurities and \cite{DOT} concerning the average conductivity for poissonian magnetic impurities), where 
as we shall see, the interacting term hapens to be non-local.\\

\hfill\break
{\it\large 3.2 Non-unitary transformation}
\hfill\break

The model beeing now defined let us consider it more explicitely. Given a magnetic field
distribution $B({\bf z})$, the static gauge field can be expressed in the Coulomb gauge (div${\bf A}=0$), using complex coordinates,
according to
\be A_z={i\over2\pi}\int d^2{\bf z}'\ {1\over\bar z-\bar z'}B({\bf z}')\ee
(recall that ${\bf z}$ means $(z,\bar z)$,  $B({\bf z})$ is not supposed to be holomorphic).
In the following we will have to distinguish between two situations,
whether, in addition to the fluctuations, there is or not a uniform magnetic field. 

When the average value of the magnetic field is non zero, then the two possible signs for the spin projection are unequivalent, 
and assuming by convention this uniform field to have the positive value $B_0$, along
the $z$ axis, the Hamiltonian takes then the two possible forms
\be H^{\pm}=-2\left(\partial_z-{1\over4\pi}\int d^2{\bf z}'\ {1\over z-z'}B({\bf z}')\right)
                \left(\partial_{\bar z}+ {1\over4\pi}  \int d^2{\bf z}'\ {1\over\bar z-\bar z'}B({\bf z}')\right) +
                (1\pm1)B({\bf z}')\ee
Where the value of $e$ and $m$ are from now on set to $1$.
Having in mind to average later on over disorder, we perform a
non-unitary transformation  on the wave-function, in order to eliminate
the quadratic term in the fluctuating magnetic field \cite{DFO1}
\be \psi({\bf z})=\ e^{\pm{1\over2\pi}\int d^2{\bf z}b({\bf z})\log\vert z-z'\vert}\ \tilde\psi({\bf z})\ee
$b=(B-B_0)/B_0$ is the dimensionless fluctuating magnetic field.
The generating function $\cal F$ which will be used later, is invariant under this transformation, as long as the Hilbert space
is not modified. This is indeed the case, qualitatively  because $b({\bf z})$  average itself to zero over the plane, consequentely
wave functions  remain square integrable under this transformation, or under the  inverse.  
At this point it is convenient to introduce the raising  and lowering  Landau level operators, 
\bea        a&=&\partial_{\bar z}+{1\over2}z\\
     a^{\dag}&=&-\partial_z+{1\over2}\bar z\eea
(the cyclotron energy is set to one)
The Hamiltonian $\tilde H^{\pm}$ acting on the wave function $\tilde\psi$ takes then the simple form
\bea \tilde H^+&=&2aa^{\dag}+{1\over\pi}\int d^2{\bf z}'{b({\bf z}')\over\bar z-\bar z'}a^{\dag}\\
     \tilde H^-&=&2a^{\dag}a-{1\over\pi}\int d^2{\bf z}'\ {b({\bf z}')\over z-z'}a\eea
This last expression, corresponding to the spin down case, makes the (expected) zero mode directly apparent. It consists of
the lowest landau itself, giving therefore a degeneracy to these particular states
in accordance with the Atiyah Singer theorem. The fact that this feature
explicitely appears here, is due to the very analogous form of the factor multiplying  $\tilde\psi$ with the Aharonov-Casher solution
\cite{ACR}.

In absence of a uniform magnetic field, considering the statistical properties of the random field, there is no priviledged orientation
given to the plane. The sign of the spin projection is therefore indifferent, and taking it negative by convention, we then perform the
same non-unitary transformation (minus sign in the exponent).
The Hamiltonian acting on $\tilde\psi$ has then the expression 
\be \tilde H=-2\partial_z\partial_{\bar z}+{1\over\pi}\int d^2{\bf z}'{b({\bf z}')\over \bar z-\bar z'}\partial_z\ee

\hfill\break
{\bf\large 4. Saddle point DOS}\\
{\it\large 4.1 Replica Supersymmetry representation}
\hfill\break

We turn now to the functionnal approach to the question of computing the d.c. transport coefficients. The objective is to average 
the generating function over disorder, which necessitates to express it in a convenient form. Although we know that im principle, 
in order to compute the mean conductivity, it is sufficient to average $\exp({\cal F})$ (the equivalent of the
partition function, instead of the free energy), we will follow the logic of this presentation, and average $\cal F$ itself.
Using the combination of the supersymmetry method of Efetov \cite{EFT} and the replica trick, it is possible to express 
$\cal F$ in a suitable form for average.\\
The supersymmetric calculus gives the possibility to view a ratio of two determinants as a superdeterminant  
\be {\det[E_F-H(\ca_z,\ca_{\bar z})+i\eta\tau_3]\over\det[E_F-H+i\eta\tau_3]}= S\det[E-H(\ca_z,\ca_{\bar z})+i\eta\tau_3]\ee
where the numerator corresponds to the grassmann sector and the denominator to the bozonic one. The sources are still ordinary 
complex fields, but are acting in the Grassmann sector. The superdeterminant is then expressed
in terms of a Gaussian integral over  superfields. 
Moreover, the replica-trick amounts  to express the logarithm according to
\be \log S\det[E-H(\ca_z,\ca_{\bar z})+i\eta\tau_3]=\lim_{q\to 0}{d\over d q}S\det[E-H(\ca_z,\ca_{\bar z})+i\eta\tau_3]^q\ee
which has to be understood by means of analytical continuation of $q$ to real values.
This allows to express $\cal F$ in the way
\be  {\cal F}[\ca_z^{\alpha},\ca_{\bar z}^{\alpha}]= \lim_{q\to 0}{d\over d q}\int\ D{\bar\Psi}D{\Psi}\ 
     e^{i\int d^2{\bf z}{\bar\Psi}\left(E-H(\ca_z,\ca_{\bar z})+i\eta\tau_3\times I_q\right){\Psi}}\ee
where the integration is performed over $2q$ components superfields, $q$ complexes and $q$ grassman components.
In this form it is now possible to average over the magnetic fluctuations. Let us consider the case with no uniform magnetic field.
The term to be averaged is 
\be <e^{-{i\over4\pi}\int d^2{\bf z}d^2{\bf z}'{\bar\Psi}({\bf z}){b({\bf z}')\over \bar z-\bar z'}D_z{\Psi}({\bf z})}>_{b}\ =\ 
     e^{{w\over2\pi}\int d^2{\bf z}_1d^2{\bf z}_2
     {\bar\Psi}({\bf z}_1)D_{z_1}{\Psi}({\bf z}_1){z_1-z_2\over \bar z_1-\bar z_2}
     {\bar\Psi}({\bf z}_2)D_{z_2}{\Psi}({\bf z}_2)}\ee
where $D_z$ replaces $\partial_z$ because of the sources and where has been utilized 
the identity
\be \int d^2{\bf z}'{1\over (\bar z_1-\bar z')(\bar z_2-\bar z')}=-\pi{z_1-z_2\over \bar z_1-\bar z_2}\ee
As already anticipated, the field theory we are obtaining, contains, in addition to the free part of the action,
\be {\cal S}_0=i\int d^2{\bf z}{\bar\Psi}({\bf z})\left(E+2D_zD_{\bar z}+i\eta\tau_3\times I_q\right)\Psi({\bf z})\ee
a quartic interacting term, which in contrary to the usual random scalar potential,
happens to be non local, because of the long range nature of the Aharonov-Bohm interaction. This term can be reexpressed using the
definition of the supertrace
\be {\cal S}_{int}={w\over2\pi}\int d^2{\bf z}_1d^2{\bf z}_2
 Str\left[{z_1-z_2\over \bar z_1-\bar z_2}D_{z_1}\Psi({\bf z}_1){\bar\Psi}({\bf z}_2)D_{z_2}
\Psi({\bf z}_2){\bar\Psi}({\bf z}_1)\right]\ee
and can nevertheless be decoupled, using a Hubbard-Strattanovich transformation, that is to say the identity
\be e^{{\cal S}_{int}}=\int D\bq DQ\ e^{-{2\pi\over w}Str\bq Q\ +\int d^2{\bf z}_1d^2{\bf z}_2
 Str\left[{z_1-z_2\over \bar z_1-\bar z_2}D_{z_1}\Psi({\bf z}_1){\bar\Psi}({\bf z}_2)Q({\bf z}_2,{\bf z}_1)+
    \bq ({\bf z}_1,{\bf z}_2)D_{z_2}
\Psi({\bf z}_2){\bar\Psi}({\bf z}_1)\right]}\ee
The functionnal integration involve  $2n\times2n$ non-local supermatrices with no particular  constraint, in contrast to the 
local theory \cite{EFT}. Permutating again the supertrace in ${\cal S}_{int}$, and integrating over the $2n$ components 
superfields, we obtain
\be\label{FIQ} {\cal F}[\ca_z^{\alpha},\ca_{\bar z}^{\alpha}]=\lim_{q\to 0}{d\over d q}\ 
          \int D\bq DQ\ e^{Str\log\left[E+2D_zD_{\bar z}+i\eta\tau_3\times I_q+i{z_1-z_2\over \bar z_1-\bar z_2}QD_z+i\bq D_z\right]
                        -{2\pi\over w}Str\bq Q}\ee
{\it\large 4.2 Saddle point approximation}
\hfill\break

The saddle point is then obtained as the stationnary  point when $Q$ and  $\bq $ are varied independantely. Expanding at 
first order the supertrace
of the logarithm in the above expression, leads to the set of equations
\bea\label{SPE} Q_0({\bf z}_1,{\bf z}_2)&=&i{w\over2\pi}<{\bf z}_1\vert\ \partial_z\ G_0\vert{\bf z}_2>\\
     \bq_0({\bf z}_1,{\bf z}_2)&=&i{w\over2\pi}\ {z_1-z_2\over \bar z_1-\bar z_2}\ <{\bf z}_1\vert\ \partial_z\ G_0\vert{\bf z}_2>\\
G_0({\bf z}_1,{\bf z}_2)&=&<{\bf z}_1\vert{1\over E+2\partial_z\partial_{\bar z}+2i\bq_0\partial_z}\vert{\bf z}_2>\eea
{\it the sources beeing set to zero for the moment}.
The main difference with the scalar case is that these relations are equalities between operators. In absence of a uniform magnetic
field, we are seeking a solution which preserves the translationnal and rotationnal invariance, meaning that $G$ has to be a function
of $p=\vert{\bf p}\vert$. In addition we look for this solution 
in a diagonal form. In that case the equation for any  diagonal components 
$g(E,p)$ of 
$G({\bf p})$ rewrites
\be g(E,p)={1\over E-{1\over2}p^2-{w\over4\pi}p^2g(E,p)}\ee
This equation admits two real solutions
\be  g^{\pm}(E,p)={2\pi\over wp^2}\left(E-{1\over2}p^2\pm\sqrt{(E-E_p^-)(E-E_p^+)}\right)\ee
 outside the energy interval $[E_p^-,E_p^+]$
 with 
\be E_p^{\pm}={1\over2}p^2\pm\sqrt{w\over\pi}p\ee
One corresponds to a stable saddle point, the other to an unstable one. By looking at the sequence
$g_{n+1}(E,p)=1/(E-{1\over2}p^2-{w\over4\pi}p^2g_{n}(E,p))$
one observe that when $E<E_p^-$, the attractive fixe point (stable) is $g^+(E,p)$, even though it is $g^-(E,p)$ when $E>E_p^+$.
This is then sufficient to analytically continue $g(E,p)$ onto the complexe plane, without any ambiguity,
\be g(Z,p)={2\pi\over wp^2}\left(Z-{1\over2}p^2-\sqrt{(Z-E_p^-)}\sqrt{(Z-E_p^+)}\right)\ee
where the square root is defined by using the convention that the argument of $Z-E_p^-$ and $Z-E_p^+$ vary in the interval $[-\pi,\pi]$.
As a result we have a branch cut situated on the segment $Z\in [E_p^-,E_p^+]$, 
\be\label{gp} g(E\pm i\eta,p)={2\pi\over wp^2}\left(E-{1\over2}p^2\mp i\sqrt{(E_p^+-E)(E-E_p^-)}\right)\ee
corresponding to a non-vanishing contribution to the density of states
\be \rho_p(E)={1\over2i\pi}\left(G_p(E-i\eta)-G_p(E+i\eta)\right)={2\over wp^2}\sqrt{(E_p^+-E)(E-E_p^-)}\ee
Therefore, the SCBA density of states is obtained by summing this last expression with the measure $pdp/2\pi$, which leads to\\
 
\be \rho(E)=\cases{0 &$E<-{w\over2\pi}$\cr
                  {1\over2w}(2E+{w\over\pi})\qquad &$-{w\over2\pi}<E<0$\cr
                           {1\over2\pi}&$E>0$}\ee
\hfill\break
and is responsible for the spontaneous breaking of the symmetry $V(2n)\to V(n)\times V(n)$, i.e. between the advanced and 
retarded sector, at the saddle-point
\be\label{SPP} G_0(E,p)={2\pi\over wp^2}\left(E-{1\over2}p^2-i\tau_3\times I_q \sqrt{(E_p^+-E)(E-E_p^-)}\right)\ee
The negative part of the DOS is actually unphysical, because the spectrum of this problem is supposed to be positive. The possible 
origin of this indesirable feature shall be discussed in the conclusive section.\\
\hfill\break
{\it\large 4.3 Broadened Landau spectrum}
\hfill\break

If we turn now to the case, where a uniform magnetic field is present, we are expecting to find a saddle-point solution
which is both invariant under magnetic translation and rotations around the origin (if we choose the symmetric gauge centered at
the origin). The representation of this group of transformation is irreducible in each Landau level, meaning that $G(E)$ has to
be decomposed on the set of projection operator $P_n$ corresponding to  each Landau Level. In other words,  $G(E)$ has to be a
function of $aa^\dag$
\be\label{gaa} G(E,aa^\dag)=\sum_{n=1}^{\infty}G_n(E)P_n\ee
For the spin-up case, the saddle point equation (with the sources set to zero again) takes the form
\be G({\bf z}_1,{\bf z}_2)=
<{\bf z}_1\vert{1\over E-2a a^\dag-{w\over\pi}{z-z'\over \bar z-\bar z'}a^\dag\ G\ a^\dag}\vert{\bf z}_2>\ee
(where   a shorthand notation has been  to express that the kernel of $a^\dag\ G\ a^\dag$ is multiplied by 
${z-z'\over \bar z-\bar z'}$).
Using the properties (see appendix) that
\be {z_1-z_2\over\bar z_1-\bar z_2}<{\bf z}_1\vert a^\dag\ G(E,aa^\dag)\ a^\dag\vert {\bf z}_2>=
<{\bf z}_1\vert a a^\dag G(E,aa^\dag)\vert {\bf z}_2>\ee
and following the lines of reasoning described in the preceeding case, we obtain (with $n$ labelling the set of Landau levels)
\be\label{g0n} G_0(E,n)=\pi^2\left(e_n(E)-i\tau_3\times I_q\rho_n(E)\right)={2\over E-2n+i\tau_3\times I_q\ \sqrt{(E-E_n^-)(E_n^+-E)}}\ee
with 
\be E_n^\pm= 2n\pm2\sqrt{{w\over\pi}n}\ee
which leads to the standard semi-circle law for the SCBA density of states\\

\be\label{dne} \rho_n(E)={1\over2\pi^2 n\alpha}\cases{\sqrt{4\alpha n-(E/b-2n)^2}\qquad&$E_n^-<E<E_n^+$\cr\cr
                     0&$otherwise$}\ee
where the parameter $\alpha={w\over b\pi}$ has been introduced, and where the dependancy with the $b$ field has been restored.
In this form it is explicitely seen that $\rho(E)$, and actually any other quantity is a function of $E/b$ and $\alpha$, the scaling 
property of the system is not broken by any UV cut-off.
The real part of the Green function is given by
\be\label{rg} e_n(E)={1\over2\pi^2 n\alpha}\cases{E/b-2n+\sqrt{(E/b-2n)^2-4n\alpha}\qquad&$E<E_n^-$\cr\cr
                   E/b-2n\qquad&$E_n^-<E<E_n^+$\cr\cr
                    E/b-2n-\sqrt{(E-2n)^2-4n\alpha}\qquad&$E>E_n^+$
                    }\ee
\hfill\break
The DOS is therefore composed of a sequence of enlarged Landau levels, with a width increasing in proportion to the
square root of the LL index $n$. The result is not very different from what is obtained for a poissonian magnetic impurities system 
\cite{DFO1,THE}
This is however different from the scalar random potential situation, where the width is not $n$ dependant, and 
this fact has a simple semi-classic interpretation \cite{AAMW,DFO2}; the semi-classic orbits in the Landau problem
consists of cyclotron orbits,
the number of time the electron circulates around such an orbite corresponds to which Landau level it belongs to.
\begin{figure}[h]
\centerline{\epsfxsize=8 cm \epsfbox{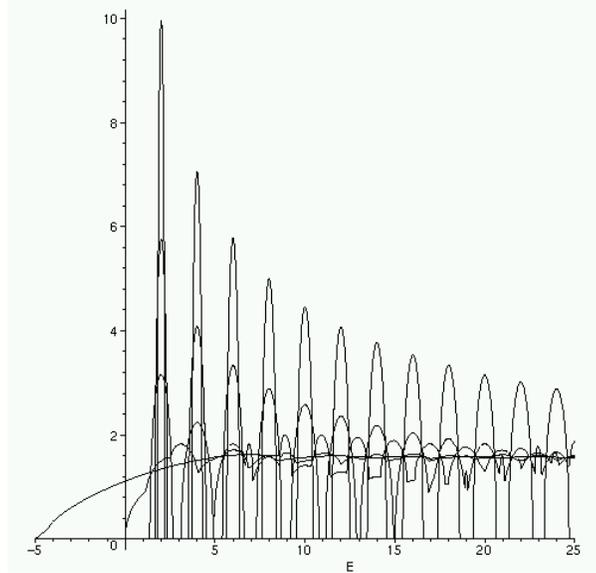}}
\caption{\small Broadening of the Landau Levels at fixed $b$ for the values $\alpha=10^{-2}, 3 10^{-2}, 10^{-1}, 1, 10$ of the disorder
parameter. 
\label{ageing}}
\end{figure}
Moreover, the 
magnetic flux passing through a given area, has fluctuations proportionnal to the square root of this area, meaning that 
the fluctuations of the flux passing through a cyclotron orbit surrounded $n$ times, is proportionnal to $\sqrt n$, giving the
typical width of the broadened Landau levels.
In the spin-down case, the saddle point equation reads
\be G({\bf z}_1,{\bf z}_2)=
<{\bf z}_1\vert{1\over E-2a^\dag a-{w\over\pi}{\bar z-\bar z'\over z-z'}a\ G\ a}\vert{\bf z}_2>\ee
and using now the property
\be { \bar z_1-\bar z_2\over z_1-z_2}<{\bf z}_1\vert a\ G(E,aa^\dag)\ a\vert {\bf z}_2>=
<{\bf z}_1\vert a^\dag a G(E,aa^\dag)\vert {\bf z}_2>\ee
we obtain the same expressions but with the modification $n\to n-1$. In particular, the lowest Landau level at $E=0$ 
remains unperturbed.\\
\hfill\break
{\bf\large 5. Saddle point Conductivity}\\
{\it\large 5.1 In absence of a uniform magnetic field}
\hfill\break

The next step is to calculate the transport coefficient at the saddle point. Using the generating function, it amounts to replace 
the integration over the $Q$ and $\bq$ matrices in (\ref{FIQ}) by the saddle-point solution $Q_0$ and $\bq_0$
of equation (\ref{SPE}). Since the saddle point doesn't break the replica symmetry, we therefore obtain the saddle point expression for 
the generating function
\be   {\cal F}_0[\ca_z^{\alpha},\ca_{\bar z}^{\alpha}]=\log 
        S\det\left[E+2D_zD_{\bar z}+2i\bq_0D_z\right]\ee
using the fact that $ {z_1-z_2\over \bar z_1-\bar z_2}Q_0({\bf z}_1,{\bf z}_2)=\bq_0({\bf z}_1,{\bf z}_2)$.
By differentiation with respect to the sources, according to receipes given in part one,  we would then obtain 
bare values for transport coefficients. However, at this point, we have to realize that the saddle point generating
function doesn't fulfil local $U(1)\times U(1)$ (gauge) invariance, namely because $Q_0$ and
$\bq_0$ are non-local operators. This originates in their dependancy with respect to 
the operators $\partial_z$ and  $\partial_{\bar z}$ 
accordingly to equation (\ref{SPE}). A straightforward procedure to restore gauge invariance would be then, to replace these operators 
by $D_z$ and $D_{\bar z}$. However, this procedure is ambiguous because $\tau_3$ does not commute with $D_z$ and $D_{\bar z}$, 
leaving any  expression of $Q_0$ to be arbitrary. Instead, if we compute the saddle point, while keeping the sources fixed,
we obtain the correct dependancy of $Q_0$ and $\bq_0$ with the sources. Let us denote $Q(\ca)$ and $\bq(\ca)$ the corresponding 
saddle point solutions. From the first part we know that it is sufficient to consider uniform sources to compute the conductivity
tensor. In that case the saddle point expression for the generating function can be written in the form
\be\label{SPF}   {\cal F}_0[\ca_z^{\alpha},\ca_{\bar z}^{\alpha}]=\log 
        S\det\left[E+\{D_z,D_{\bar z}\}+i\bq(\ca) D_z-iD_{\bar z}Q(\ca)\right]-{2\pi\over w}Str \bq(\ca)Q(\ca)\ee
with the saddle point constraint
\bea Q(\ca)&=&i{w\over2\pi}D_z\ G(\ca)\\
     \bq(\ca)&=&-i{w\over2\pi}\ G(\ca) D_{\bar z}\\
 \label{SP3} G(\ca)&=&{1\over E+\{D_z,D_{\bar z}\}+{w\over2\pi}\ \{G(\ca),D_{\bar z}D_z\}}\eea
The computation of transport coefficients necessitates then to solve this last equation. Assuming that $G(\ca)$ is analytic with 
respect to the sources, it is then sufficient to solve at first order in the source. Consider the expansion
\be G(\ca)=G_0+G_\alpha^z \ca_z^\alpha+G_\alpha^{\bar z} \ca_{\bar z}^\alpha+..\ee
Deriving (\ref{SPF}) twice with respect to the sources, according to the saddle point constraints leads to the following expression 
for the conductivity coefficients
\bea\label{sxx} \sigma_{xx}&=&{1\over4\pi V} Tr\left\{G_0+
{w\over4\pi}G_0^2-2\left[v_{\bar z}+{w\over4\pi}(v_{\bar z}G_0+G_0^{\dag}v_{\bar z})\right]G_1^z\tau_1\right\}\\
\label{sxy}\sigma_{xy}^{I}&=&{i\over2\pi V} Tr\left\{
{w\over4\pi}G_0^2-2\left[v_{\bar z}+{w\over4\pi}(v_{\bar z}G_0+G_0^{\dag}v_{\bar z})\right]G_1^z\tau_1\right\}\tau_3\eea
where the trace is yet restricted to the Grassmann sector, and $G_0^{\dag}$ on this sector reads
\be G_0^{\dag}=\pmatrix{G_0^-&0\cr 0&G_0^+\cr}\ee
whereas
\be G_0=\pmatrix{G_0^+&0\cr 0&G_0^-\cr}\ee
In absence of a net magnetic field, we expect $G_\alpha^z$ to be a  function of $p$ and $\bar p$ because of translationnal 
invariance. Solving (\ref{SP3}) with this assumption we obtain
\be G_\alpha^z({\bf p})=-{\bar p\over2}\ {\left(1+{w\over4\pi}(G_0+G_0^{\dag})\right)G_0\ G_0^{\dag}
                           \over1-{w\over4\pi}p^2\ G_0G_0^{\dag}}\ \tau_\alpha\qquad\qquad\alpha=1,2\ee
the striking feature of this last expression is that it is not defined when $p$ is such that $E_p^-<E_F<E_p^+$, the denominator vanishes
because in this interval we have (\ref{gp}) 
\be G_0^+G_0^-={4\pi\over w p^2}\ee
The reason for this is that the procedure of keeping sources in the computation of the saddle point, amounts to incorporate
ladder diagrams contribution, when the trace is taken. These diagrams are in principle responsible for weak localization corrections,
and since the system, because of the magnetic fluctuations, is not time-reversal invariant, they are actually the only ones.
Therefore this divergency indicates that $G(\ca)$ is not an analytic fonction with respect to  the sources, which has to be related
to the eventuality that the system is localized.
The expression of $G_\alpha^z({\bf p})$ suggests the possibility of a self-consistent       
treatment of the problem \cite{VWF};
this would however  require a more complete treatement, 
beyond saddle point approximation, which is out of the scope of this paper.\\
\hfill\break
{\it\large 5.2 Conductivity in presence of a uniform magnetic field}
\hfill\break

The situation is quite different when an external uniform magnetic field is added to the problem. In that case, the IR divergency
observed preceedingly happens to be removed. Indeed, the solution to (\ref{SP3}) can be searched in the form
\be G_\alpha^z(a,a^{\dag})=a^{\dag}g(aa^{\dag})\tau_\alpha\ee
and using the rules (for an arbitrary function $f$),
\bea  af(a^{\dag}a)&=&f(aa^{\dag})a\\
      a^{\dag}f(aa^{\dag})&=&f(a^{\dag}a)a^{\dag}\eea
we obtain the solution
\be G_\alpha^z(a,a^{\dag})=-i{\left[1+{w\over4\pi}\left(G_0(aa^{\dag})+G_0^{\dag}(a^{\dag}a)\right)
\right]G_0(aa^{\dag})\ G_0^{\dag}(a^{\dag}a)
                           \over1-{w\over2\pi}(2aa^{\dag}-1)\ G_0(aa^{\dag})\ G_0^{\dag}(a^{\dag}a)}
                           \ a^\dag\ \tau_\alpha\ee
for $\alpha=1,2$ and 
\be G_3^z(a,a^{\dag})=-i{\left[1+{w\over4\pi}\left(G_0(aa^{\dag})+G_0(a^{\dag}a)\right)
\right]G_0(aa^{\dag})\ G_0(a^{\dag}a)
                           \over1-{w\over2\pi}(2aa^{\dag}-1)\ G_0(aa^{\dag})\ G_0(a^{\dag}a)}
                           \ a^\dag\ \tau_3\ee
\begin{figure}[h]
\centerline{{\epsfxsize=7 cm \epsfbox{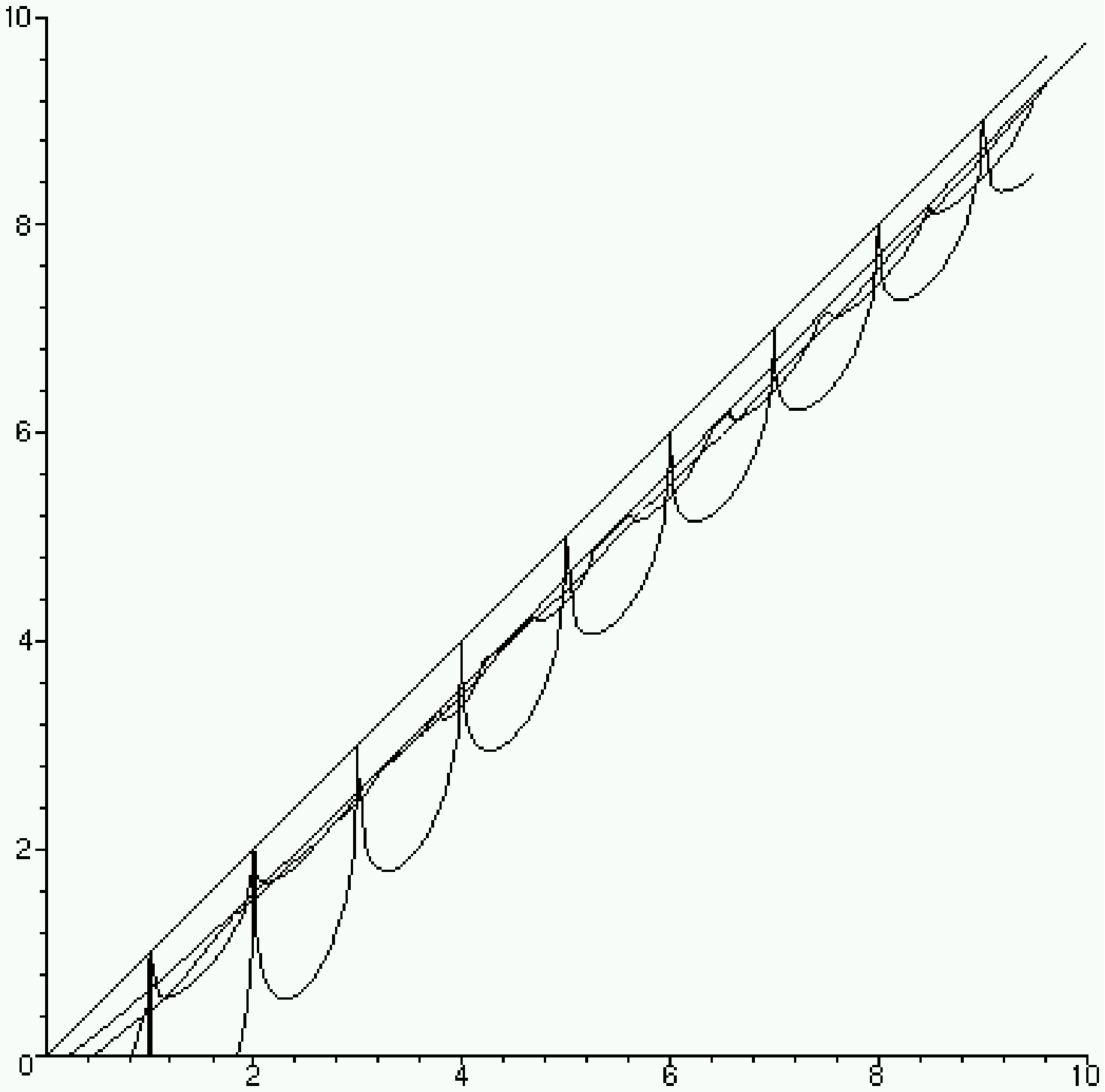}} {\epsfxsize=7 cm \epsfbox{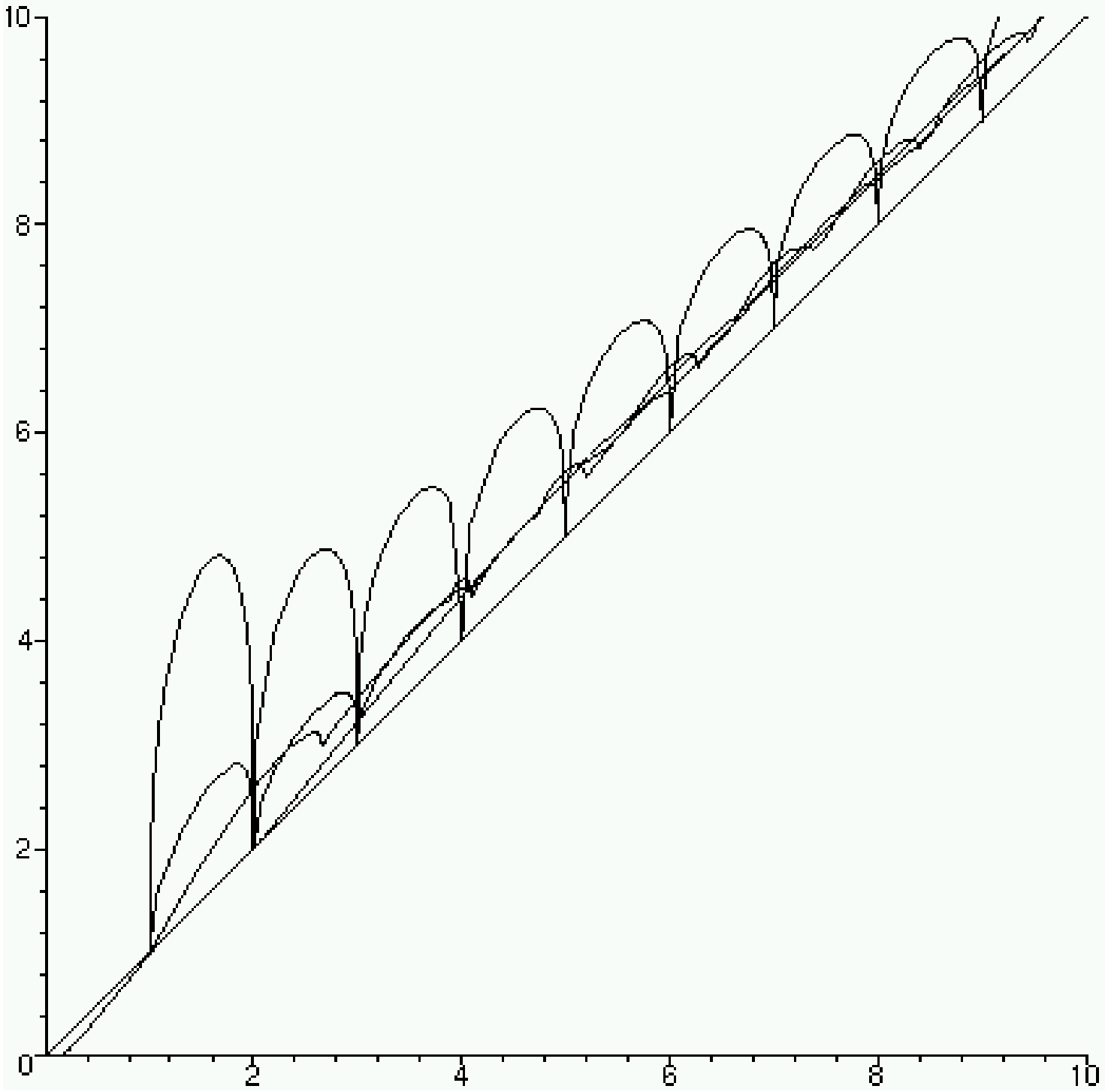}}}
\caption{\small Transverse conductivity in unit of ${e^2\over h }$ for the spin-up (left) and spin-down (right) cases
 for $\alpha=10^{-2}, 10^{-1}, 1, 10$, compared
to the classical value (straight line). The $x$ axes is the Landau level filling number.
\label{ageing}}
\end{figure}
where $G_0(aa^{\dag})$ is defined according to (\ref{gaa}) and (\ref{g0n}). Inserting this last expression in (\ref{sxx}) and (\ref{sxy})
allows to compute the conductivity tensor. However for $\sigma_{xx}$ it is convenient to use the following Ward identity 
which is a consequence of global $U(1)\times U(1)$ invariance,
\be Tr\left\{G_0+
{w\over4\pi}G_0^2-2\left[v_{\bar z}+{w\over4\pi}(v_{\bar z}G_0+G_0 v_{\bar z})\right]G_3^z\tau_3\right\}=0\ee
and which is the saddle point version of (\ref{GWI}). Defining
\be g_n^s=\pi^2\left(e_n(E_F)-is\rho_n(E_F)\right)\ee
with $s=\pm$ and  $e_n(E_F)$ and $\rho_n(E_F)$ given by (\ref{rg}) and (\ref{dne})
we can then express $\sigma_{xx}$ in the form (recall $\alpha={w\over b\pi}$)\\
\be\label{sxx} \sigma_{xx}=-{1\over2\pi^2}\sum_{s,s'}\sum_{n=1}^\infty\ ss'(n-1){\left(1+{\alpha\over4}(g_n^s+g_{n-1}^{s'})\right)^2\ 
                 g_n^sg_{n-1}^{s'}\over1-{\alpha\over2}(2n-1)g_n^sg_{n-1}^{s'}}\ee
\hfill\break
In the strong magnetic field limit, i.e.  when the Landau Levels do not overlap, it is quite easy to see that  $\sigma_{xx}$ vanishes.
Surprisingly enough this hapens to be true whatever $b$ is (numerical check)
\be \sigma_{xx}=0\ee
Concerning the transverse conductivity, we obtain for the dissipative part
\be\label{sxy} \sigma_{xy}^I=\sum_{n=1}^\infty\left\{ {1\over4n}(E_F/b-2n)\rho_n(E_F)+{i\over\pi^2}
                 \sum_{s=\pm}\ s(n-1){\left(1+{\alpha\over4}(g_n^s+g_{n-1}^{-s})\right)^2\ 
                 g_n^sg_{n-1}^{-s}\over1-{\alpha\over2}(2n-1)g_n^sg_{n-1}^{-s}}\right\}\ee
although the contribution comming from the magnetization is obtain by using the Streda formula (\ref{std}) 
\be \sigma_{xy}^{II}={1\over b}N(E_F)-\sum_{n=1}^\infty{1\over2}(E_F/b+2n)\rho_n(E_F)\ee
Altogether the sum simplifies exactly to
\be \sigma_{xy}^{up}={1\over b}N(E_F)-\sum_{n=1}^\infty\rho_n(E_F)={1\over b}N(E_F)-\rho(E_F)\ee
For  the spin-down case, all the $n$ appearing in the fraction of expression (\ref{sxx}) and (\ref{sxy}) are shifted by $-1$.
The conclusion concerning $\sigma_{xx}$ is not modified, the longitudinal conductivity vanishes exactly for all values of the
magnetic field. For $\sigma_{xy}$ we obtain a slightly modified expression, exept for the sign relative to the ``classical'' value,
\be \sigma_{xy}^{down}={1\over b}N(E_F)+\sum_{n=1}^\infty{E_F\over2bn}\rho_n(E_F)\ee
which again is obtained with no approximation. The Hall conductance  is plotted in figure $2$ for the two orientations of the spin.
In the strong magnetic field limit we obtain oscillations of the transverse conductivity, in qualitative agreement with the
conclusion of reference \cite{DOT} for the spin-down case.
The spin-up plot ressembles qualitatively the numerical results obtained in reference \cite{NHD}. 
\begin{figure}[h]
\centerline{\epsfxsize=7 cm \epsfbox{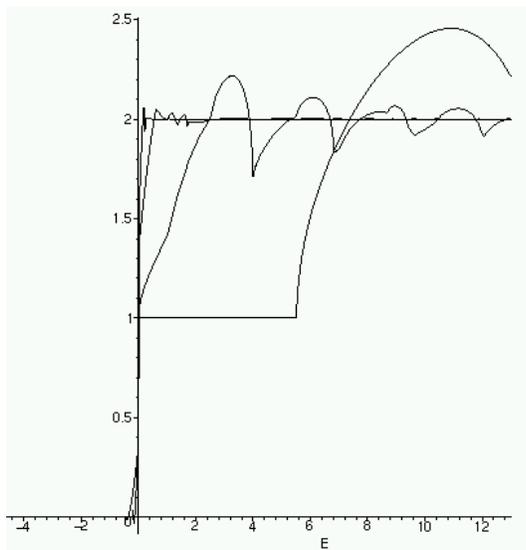}}
\caption{\small Difference between the spin-up and the spin-down transverse conductivity in unit of ${e^2\over h}$ 
for $b=5, 1, .1,.01$ and $w=\pi$, versus Fermi energy.
\label{ageing}}
\end{figure}

The limit of vanishing mean magnetic field is a bit more mysterious. It should be first noticed that $\sigma_{xy}$ 
as a function of electronic integrated density is divergent. Although paradoxical, one can consider this to be 
coherent with the fact that $\sigma_{xx}$ vanishes, if one think of the semi-classical relation between 
these two coefficients \cite{AFS}. Anyway, something interesting happens when we compare the Hall conductance for the two 
different spin projections in this limit, as a function of the Fermi energy (see figure $3$). This difference takes the quantized
value of $2$ (in unit of ${e^2\over h}$) when $b$ goes to zero. The first quantum is simply traced back to the additionnal Landau
level lying at $E=0$ in the spin-down situation and which is not affected by disorder (see the strong magnetic field curve $b=5$ in
figure $3$). The second one seems to indicate that the system restores the chiral symetry (considering spin-up and spin down situations
as a whole) when the average $b$ goes to zero, 
by reintroducing progressively a zero energy state in the spin-up spectrum. In addition this state restores parity, by contributing
with the opposite sign to the Hall conductance.  
In this sense, the limit of zero magnetic field is unstable with respect to spin-flip at $E=0$.

\hfill\break
\hfill\break
{\bf\large 6. Conclusion}
\hfill\break

The formalism presented in this paper, which proposes to compute directly 
static transport coefficients from a generating function, has been
used at the basic level of the saddle point approximation. Already at this stage was exemplified the possibility of incorporating 
first dominant contributions, namely the ladder diagrams contribution, by keeping uniform sources when computing the saddle point. 
In the exemple we considered, the white noise random magnetic field, it was observed that in absence of an external uniform
magnetic field, the saddle point solution is not analytic with respect to the sources indicating a non-simply diffusive behavior.
Slowly spatially varying sources should be considered in order to analyse further this question, and a self-consistent argument
analogous to the one of Vollhardt and  W\"olfle for the scalar disorder case \cite{VWF} could be elaborated. Furthermore, this formalism
should be possibly generalized for the computation of dynamical transport coefficients, and eventually extended to 
the non-linear regime. 

Concerning the random magnetic field problem, the free parameter $\lambda$ which was introduced in section $3$,
was only considered for two particular values, which presumably insures the system to remain scale invariant quantum mechanically.
The fact that the longitudinal conductivity was suprizingly found to remain exactly zero, when Landau levels do  overlap, although
not really  conclusive (within the saddle-point approximation), 
is in favour of this hypothesis. Interestingly enough, although may be more technically involved, 
would be to consider this model for arbitrary $\lambda$, in particular to see wether it is possible to find at this level 
of approximation a relation between $\sigma_{xx}$ and $\sigma_{xy}$ analogous to the one obtained by Ando et al. for the 
scalar random potential \cite{AFS}. In addition a flow between these two points under scale transformation might be expected. Beforehand 
the problem of undesirable negative energy states which appear in zero magnetic field has to be cured, may be by avoiding the
non-unitary transformation, which presumably causes some non-normalizable states to re-enter the Hilbert space. This could possibly
clarify the nature of the zero energy state in the averge magnetic field limit, 
which, as suggested by the analyses of figure $3$ seems to spontaneously break
parity and to be able to carry one quantum of Hall current.\\
\hfill\break
\hfill\break 
 {\bf Acknowledgements}\\ 
 I am pleased to thank J. Desbois, S. Ouvry and C. Texier for useful discussions concerning their approach to this problem,
 J.M. Leinaas, and H.A. Weidenmueller for discussing the supersymmetric approach to the transport problem.
 I am thankful   to  the Institute of physics of the Oslo University for hospitality, and 
 the National Reasearch Council of Norway for financial support.

\hfill\break
{\bf\large Appendix}
\hfill\break

In this section we justify the use of the formula
\be {z_1-z_2\over\bar z_1-\bar z_2}<{\bf z}_1\vert\ a^\dag\ f(aa^\dag)\vert {\bf z}_2>=
<{\bf z}_1\vert\ f(aa^\dag)\ a\ \vert {\bf z}_2>\ee
or equivalentely
\be {z_1-z_2\over\bar z_1-\bar z_2}<{\bf z}_1\vert\ f(aa^\dag)\ a^\dag\ \vert {\bf z}_2>=<{\bf z}_1\vert\ a f(aa^\dag)\ \vert {\bf z}_2>\ee
where $f$ is an arbitrary function of $aa^\dag$, which is actually defined by the values $f_n$ it takes on each Landau Level
\be f(aa^\dag)=\sum_{n=1}^\infty f_n P_n\ee
To this end, it is convenient to introduce the raising and lowering operators of angular momentum
\bea 
     b&=&\partial_z+{1\over2}\bar z\\
     b^\dag&=&-\partial_{\bar z}+{1\over2}z\eea
and a generating function for the set of projection operators on the Landau Levels, namely the thermal propagator
\be G_t=\ e^{-aa^\dag t}=\sum_{n=1}^\infty e^{-nt} P_n\ee
Then, according to the commutation rules between the $a$, $a^\dag$ and $b$, $b^\dag$  operators we have the property
\be [\hat z,G_t]= [a+b^\dag,G_t]=[a,G_t]=(1-e^t)aG_t\ee
as a consequence, it turns out that 
\be <{\bf z}_1\vert\ a G_t\ \vert {\bf z}_2>={e^{-t}\over1-e^{-t}}\ (z_1-z_2)<{\bf z}_1\vert\ G_t\ \vert {\bf z}_2>\ee
In the same way we have
\be <{\bf z}_1\vert\ a^\dag G_t\ \vert {\bf z}_2>={1\over1-e^{-t}}\ (\bar z_1-\bar z_2)<{\bf z}_1\vert\ G_t\ \vert {\bf z}_2>\ee
which leads to obtain
\be {z_1-z_2\over\bar z_1-\bar z_2}<{\bf z}_1\vert\ a^\dag\ G_t\ \vert {\bf z}_2>=
<{\bf z}_1\vert\ G_t\ a\ \vert {\bf z}_2>\ee
or equivalentely
\be {z_1-z_2\over\bar z_1-\bar z_2}<{\bf z}_1\vert\ G_t\ a^\dag\ \vert {\bf z}_2>=
<{\bf z}_1\vert\ a\ G_t\ \vert {\bf z}_2>\ee
$G_t$ beeing a generating function for the Landau Levels projection operator, the formula is therefore valid for an arbitrary 
function of $aa^\dag$ as stated above.

\end{document}